# Tutorial on Silicon Photonics Integrated Platform Fiber Edge Coupling

Sergey S. Avdeev[1,2], Aleksandr S. Baburin[1,2], Evgeniy V. Sergeev[1], Alexei B. Kramarenko[1], Arseniy V. Belyaev[1], Danil V. Kushnev[1], Kirill A. Buzaverov[1], Ilya A. Stepanov[1], Vladimir V. Echeistov[1,2], Sergey V. Bukatin[1], Ali Sh. Amiraslanov[1], Evgeniy S. Lotkov[1], Dmitriy A. Baklykov[1], and Ilya A. Rodionov[1,2]*

[1] Shukhov Labs, Quantum Park, Bauman Moscow State Technical University, Moscow, 105005, Russia
[2] Dukhov Automatics Research Institute, VNIIA, Moscow 127030, Russia
*e-mail: irodionov@bmstu.ru

**ABSTRACT:** Photonic integrated circuits (PICs) play a crucial role in almost every aspect of modern life, such as data storage, telecommunications, medical diagnostics, green energy, autonomous driving, agriculture, and high-performance computing. To fully harness their benefits, an efficient coupling mechanism is required to successfully launch light into waveguides from fibers. This study introduces low-loss coupling strategies and their implementation for a silicon nitride integrated platform. Here we present an overview of coupling technologies, optimized designs, and a tutorial on manufacturing techniques for inverted tapers, which enable effective coupling for both transverse-magnetic and transverse-electric modes. The optimized coupling losses for the UHNA-7 fiber and the inverted taper $Si_3N_4$ coupler reached ∼-0.15 dB at 1550 nm per connection for single-mode waveguides with 220×1200 nm cross section. The measured coupling losses in the inverted taper coupler with a standard single-mode fiber were ∼-1.50 dB at 1550 nm per connection for the same platform.

## INTRODUCTION

With the growing demand for high-speed and compact devices, photonic integrated circuits (PICs) are attracting significant interest because of their high bandwidth and compatibility with large-scale integration technologies. Silicon nitride ($Si_3N_4$) is the ideal platform for photonic integrated circuit applications such as LiDAR [1], optical sensing [2–7], bio-spectroscopy [8], communication [9] and for uses in the quantum domain, such as quantum key distribution (QKD), quantum computing [10,11] and quantum sensing [12]. In recent years, $Si_3N_4$ has attracted significant attention as a PIC platform because of its low propagation loss, compatibility with heterogeneous integration, extended transparent bandwidth, and lower susceptibility to errors during the lithography and etching processes compared with the silicon-on-insulator (SOI) platform [13–16]. Typically, PICs require an effective coupling mechanism to launch light into the waveguide from a fiber. To this end, grating and edge couplers are widely employed [17–43]. Grating couplers are used in out-of-plane coupling. For high-volume manufacturing by standard production techniques, for example, in MPW fabs, gratings exhibit a rather low coupling efficiency of ∼-5.0 dB and inherently suffer from limited wavelength bandwidth [17–23]. To enhance the coupling efficiency, an inverted taper could be introduced, as it demonstrates a coupling loss of down to ∼-0.5 dB [23–26]. However, it should be noted that waveguide type and dimensions and fiber/waveguide mode size differences play a crucial role in coupling efficiency.

Edge couplers exhibit high in-plane light coupling efficiency and a broader spectral bandwidth. Most edge couplers are designed to entail long adiabatic tapers with lengths exceeding several hundred micrometers, thereby enhancing the mode transfer efficiency. Several techniques make it possible to further improve coupling efficiency, like the 2×2 couplers [29–33], mode multiplexers [20,33–35], and polarization splitters/rotators [36–43]. The compact form of the adiabatic taper drastically facilitates its applicability in high-density PICs [17]. To obtain a short and efficient taper that is applicable in high-density PICs, various taper configurations, including nonlinear, stepwise cascaded, multilayered, and metamaterials, as well as double-tip and sinusoidal taper, have been extensively analyzed for various photonic platforms [21–29]. However, unlike other platforms, compact tapers for $Si_3N_4$ platform edge couplers have rarely been investigated or reported. The main published results are presented in Table 1.

**Table 1.** Si$_3$N$_4$ taper coupling losses overview

| Reference / Wavelength | Taper length / width | Coupling efficiency TE mode, Losses | 3-dB alignment tolerance Horizontal | 3-dB alignment tolerance Vertical | Fiber type (mode size) | Waveguide crossection / technology |
|---|---|---|---|---|---|---|
| [44] Calculated 1550 nm | 500 μm / 300 nm | -0.087 dB | ±3.6 μm | ±3.5 μm | UHNA-3 (4.1±0.3 μm) | 100×900 nm / a single step litho |
| [44] Measured 1550 nm | 500 μm / 300 nm | -0.17 dB | ±3.8 μm | ±3.6 μm | UHNA-3 (4.1±0.3 μm) | 100×900 nm / a single step litho |
| [45] Measured 1550 nm | 500 μm / 180 nm | -2.0 dB | - | - | LPMFs (2.5±0.3 μm) | 400×700 nm / polished with diamond films |
| [46] Calculated 1550 nm | 45 μm / 750 nm | -0.58 dB | ±1.0 μm | ±1.0 μm | UHNA-3 (4.1±0.3 μm) | 300×1000 nm / shadow mask |
| [46] Measured 1550 nm | 45 μm / 750 nm | -1.47 dB | - | - | UHNA-3 (4.1±0.3 μm) | 300×1000 nm / shadow mask |
| [47] Calculated 1550 nm | 76 μm / 200 nm | -0.29 dB | ±2.0 μm | ±2.0 μm | UHNA-3 (4.1±0.3 μm) | 500×2000 nm / a single step litho |
| [47] Measured 1550 nm | 76 μm / 200 nm | -0.36 dB | ±3.5 μm | ±3.3 μm | UHNA-3 (4.1±0.3 μm) | 500×2000 nm / a single step litho |
| Current work Calculated 1550 nm | 360 μm / 275 nm | -2.52 dB | ± 3.2 μm | ± 2.9 μm | SMF-28 (10.5±0.5 μm) | 220×1200 nm / double step litho, 5 μm oxide thickness |
| Current work Calculated 1550 nm | 360 μm / 275 nm | -1.55 dB | ± 2.0 μm | ± 1.5 μm | SM1500es (4.0±0.5 um μm) | 220×1200 nm / double step litho, 5 μm oxide thickness |
| Current work Calculated 1550 nm | 360 μm / 275 nm | -0.75 dB | ± 1.7 μm | ± 1.3 μm | UHNA-7 (3.2±0.3 μm) | 220×1200 nm / double step litho, 5 μm oxide thickness |
| Current work Calculated 1550 nm | 360 μm / 275 nm | -0.55 dB | ± 1.5 μm | ± 1.2 μm | LPMFs (2.5±0.3 μm) | 220×1200 nm / double step litho, 5 μm oxide thickness |
| Current work Measured 1550 nm | 360 μm / 275 nm | -3.28 dB | ± 2.8 μm | ± 2.4 μm | SMF-28 (10.5±0.5 μm) | 220×1200 nm / double step litho, 5 μm oxide thickness |
| Current work Measured 1550 nm | 360 μm / 275 nm | -0.97 dB | ± 1.4 μm | ± 1.1 μm | SM1500es (4±0.5 um μm) | 220×1200 nm / double step litho, 5 μm oxide thickness |
| Current work Measured 1550 nm | 360 μm / 275 nm | -0.81 dB | ± 1.2 μm | ± 0.8 μm | UHNA-7 (3.2±0.3 μm) | 220×1200 nm / double step litho, 5 μm oxide thickness |
| Current work Measured 1550 nm | 360 μm / 275 nm | -1.25 dB | ± 0.9 μm | ± 0.7 μm | LPMFs (2.5±0.3 μm) | 220×1200 nm / double step litho, 5 μm oxide thickness |
| Current work Calculated Multi-tip taper 1550 nm | 265 μm / 8.55 μm | -0.51 dB | ± 5 μm | ± 3 μm | SMF-28 (10.5±0.5 μm) | 220×1200 nm / double step litho, 8 μm oxide thickness |
| Current work Measured Multi-tip taper 1550 nm | 265 μm / 8.55 μm | -1.50 dB | - | - | SMF-28 (10.5±0.5 μm) | 220×1200 nm / double step litho, 8 μm oxide thickness |
| Current work Calculated 1550 nm | 280 μm / 500 nm | -0.10 dB | ± 1.2 μm | ± 0.9 μm | UHNA-7 (3.2±0.3 μm) | 220×1200 nm / double step litho, 8 μm oxide thickness |
| Current work Measured 1550 nm | 280 μm / 500 nm | -0.15 dB | - | - | UHNA-7 (3.2±0.3 μm) | 220×1200 nm / double step litho, 8 μm oxide thickness |

Currently, the best achieved efficiency of -0.17 dB was demonstrated by the adiabatic Si3N4 coupler based on a taper with a length of 500 μm for UHNA-3 fiber/100×900 nm waveguide interface, cleaved with a good cleaving position tolerance [37].

This paper serves as a tutorial on finding the best repeatable wafer-scale compatible coupling strategy for the Si$_3$N$_4$

platform. The entire Si3N4 waveguide fabrication comprises waveguide fabrication and plasma dicing based on deep silicon etching, which provides optical quality chip edge forming. In this article a universal approach for taper design is for standard fiber/chip coupling are calculated to be ∼-0.51 dB and measured to be ∼-1.50 dB for 1550 nm per connection.

## COUPLING STRATEGIES

**Tapered waveguide simulation.** For PICs, low-loss edge coupling special waveguide structures, which serve as mode size converters, are used [48,49]. The optical mode size and shape change during propagation through the tapered waveguide to achieve higher coupling efficiency between two modes with different cross-sections. They are designed to operate adiabatically: the waveguide local first-order mode should propagate through the tapered waveguide while undergoing relatively little mode conversion compared with the higher-order modes or radiation modes. This adiabatic operation is realized in the taper design by gradually increasing the taper cross-section size and decreasing the mode size from typical diameter of 5-10 μm in fiber to the order of several microns in the waveguide. Different designs of adiabatic tapers have been proposed for Si3N4 platforms, including linear [48], exponential [22], parabolic [50], and multi-sectional tapers [24,25]. Most often, a linear waveguide tapers were used. In this linear adiabatic taper mode, conversion occurs more easily in the wider portion [22].

For low-loss transmission, the linear taper can be extended to be long enough to ensure adiabatic propagation. However, a longer taper could lead to higher propagation losses. This problem becomes more critical when the two modes connected by the linear taper possess higher cross-sectional differences. In this case, the taper starts with a width that is much narrower than one of the waveguide. This makes the taper wall roughness' influence on losses much more noticeable. Thus, the taper length should be reduced to ensure the lowest possible losses and to reduce the device footprint. Modeling started with the waveguide geometry determination to ensure on fundamental mode excitation (Figure 2a). Modeling was carried out in Ansys Lumerical FDE, on the basis of which a waveguide width of 1200 nm was chosen. To determine the optimal taper geometry, modeling was performed in Ansys Lumerical FDTD based on the numerical solution of Maxwell's equations. The taper 3D model is shown in Figure 1a.

The length of the fabricated tapers varied from 200 μm to 520 μm, (Figure 2 (c-f)). In turn, the width at the beginning of the taper varied from 50 to 500 nm with a waveguide width of 1200 nm. The smallest feature size is determined by the technological capabilities of electron lithography technology. The graph shows the dependance obtained during the simulation for the different optical fibers. In this study, the taper length was 360 μm for high transmission maintaining a relatively small device footprint. To standardize the technological process, a taper geometry (width 280 nm, length 360 μm) was chosen to ensure maximum coupling efficiency for all types of optical fibers. The light was emitted from the fiber onto the mode size converter, as shown in the top-down view in Figure 2h.

presented, which was used to simulate several types of inverted taper couplers. The coupling losses per coupler for the UHNA-7 fiber/chip coupling are calculated to be ∼-0.75 dB and measured to be ∼-0.81 dB for 1550 nm per connection.

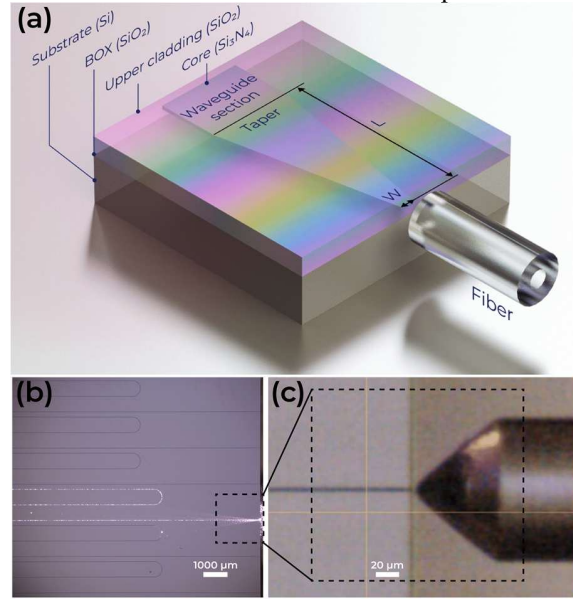

**Figure 1.** 3D visualization of coupled taper with fiber (a), Light propagation through the waveguide (b), Coupling of lensed fiber with chip facet (c).

**Edge facet.** To provide high taper coupling efficiency, the PIC dies are typically polished to obtain high optical grade quality edges [51]. Although this process works well with small chip sizes, it's scaling up to wafer level is impractical. There is also used the process of silicon dioxide thick layers wet etching, but it is anisotropic, that negatively affects the quality of the optical facets [52]. Other solutions for PICs optical facets fabrication need to be investigated. A promising technology is the reactive plasma dicing [53].

We previously demonstrated the lowest propagation losses currently available in the near-infrared wavelength (as low as 0.55 dB/cm) in single-mode Si3N4 submicron waveguides (220×550 nm) [54]. Our study continues the previous research and presents a plasma-based process for optical grade edge facets fabrication. The fabrication process of low-loss silicon nitride photonic integrated circuits with light coupling through edge couplers is shown in Figure 3. The presented technology is used to dice a chip, that consists of 220-nm-thick stoichiometric low-pressure chemical vapor Si3N4 layer grown on the 525-μm silicon substrate oxidized with the 2.5-μm thickness and covered with 2.5-μm oxide. A standard sequence of basic operations is involved for fabrication (Figure 3). First, the waveguide structure alignment markers were patterned using electron-beam lithography. The pattern is then transferred to the Si3N4 device layer by reactive ion plasma etching (RIE) [55]. After nanotopology fabrication, the chip is plasma diced. The dicing process consists of the following stages: surface preparation, photoresist spin coating, photolithography, photoresist development, SiO2-Si3N4-SiO2 thin film stack etching, photoresist removal, and silicon etching by the Bosch process [55].

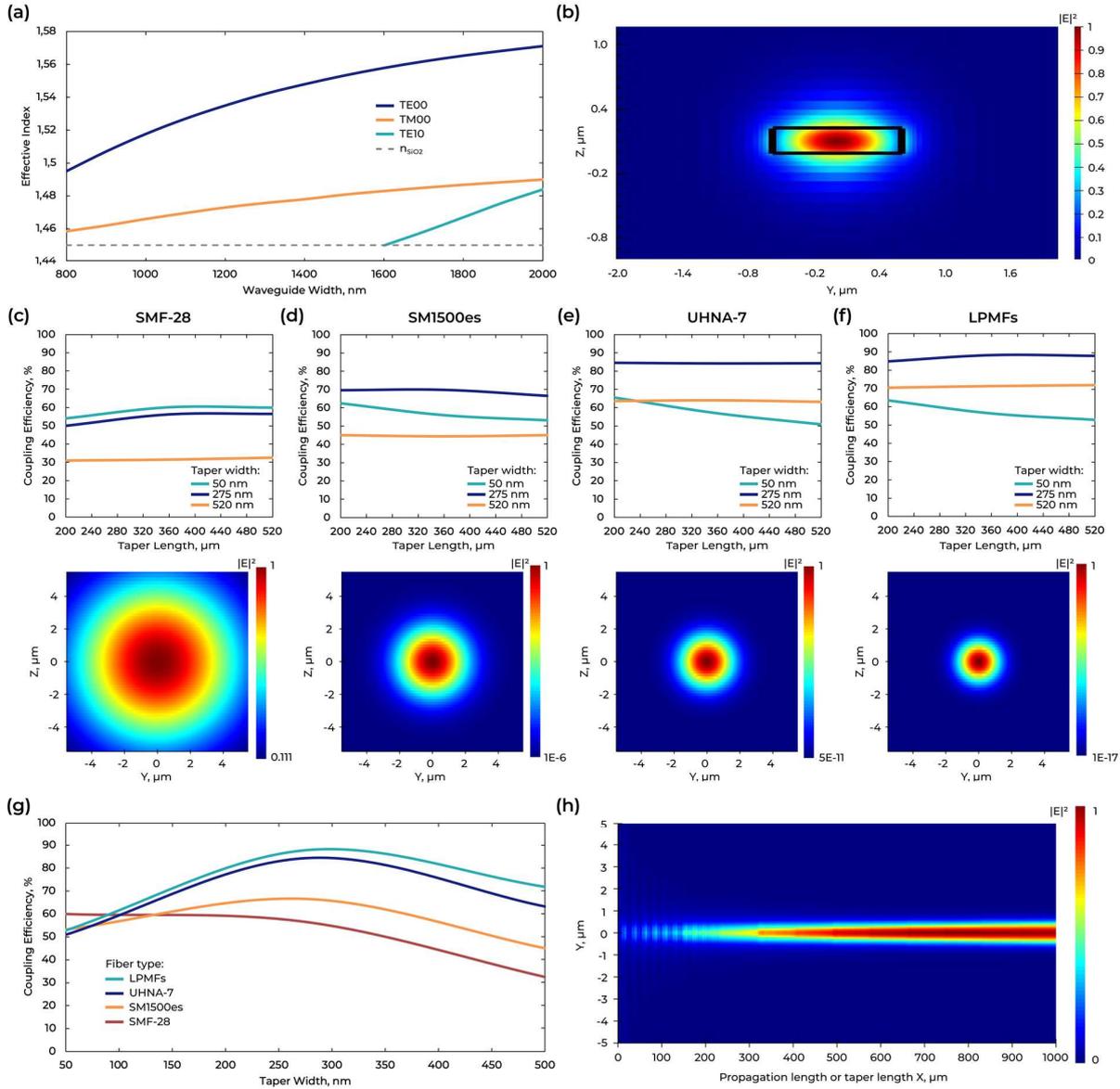

**Figure 2.** Calculation of $Si_3N_4$ effective index of modes excited in a waveguide as a function of waveguide width (a), Fundamental mode excited in the waveguide with 220×1200 nm cross section (b), Calculation of the taper length at different width for different fiber types with wavelength 1550 nm (c-f), Calculation of the taper width with fixed length of taper (360 um) for different fiber types with wavelength 1550 nm (g) and simulation of mode propagation for lensed fiber (h).

To pattern the silicon dioxide layer, 5 μm thick SPR220 photoresist was used. Patterning processes were performed using laser lithography. The topology was then transferred to the multilayer stack ($SiO_2$-$Si_3N_4$-$SiO_2$) with reactive ion etching. The next step was silicon wafer dicing using the Bosch process. Optical microscopy and scanning electron microscopy were used to characterize the samples and control process quality.

The mentioned above SPR220 positive photoresist pattern transfer mask with an anti-reflective layer using laser lithography should have a vertical profile to be further used for vertical edge etching. It lets avoiding the effect of refraction during light coupling from the optical fibers to the chip. Resist thickness was chosen based on plasma etching process selectivity to ensure etching of $SiO_2$-$Si_3N_4$-$SiO_2$ layers with a total thickness of 5.2 μm. The laser lithography dose and focus were varied to obtain the pattern mask vertical profile. for the structural vertical angle exposure in thick resist layers, multi-pass lithography was used. In this study, four- and six- pass algorithms were applied (Figure 4b).

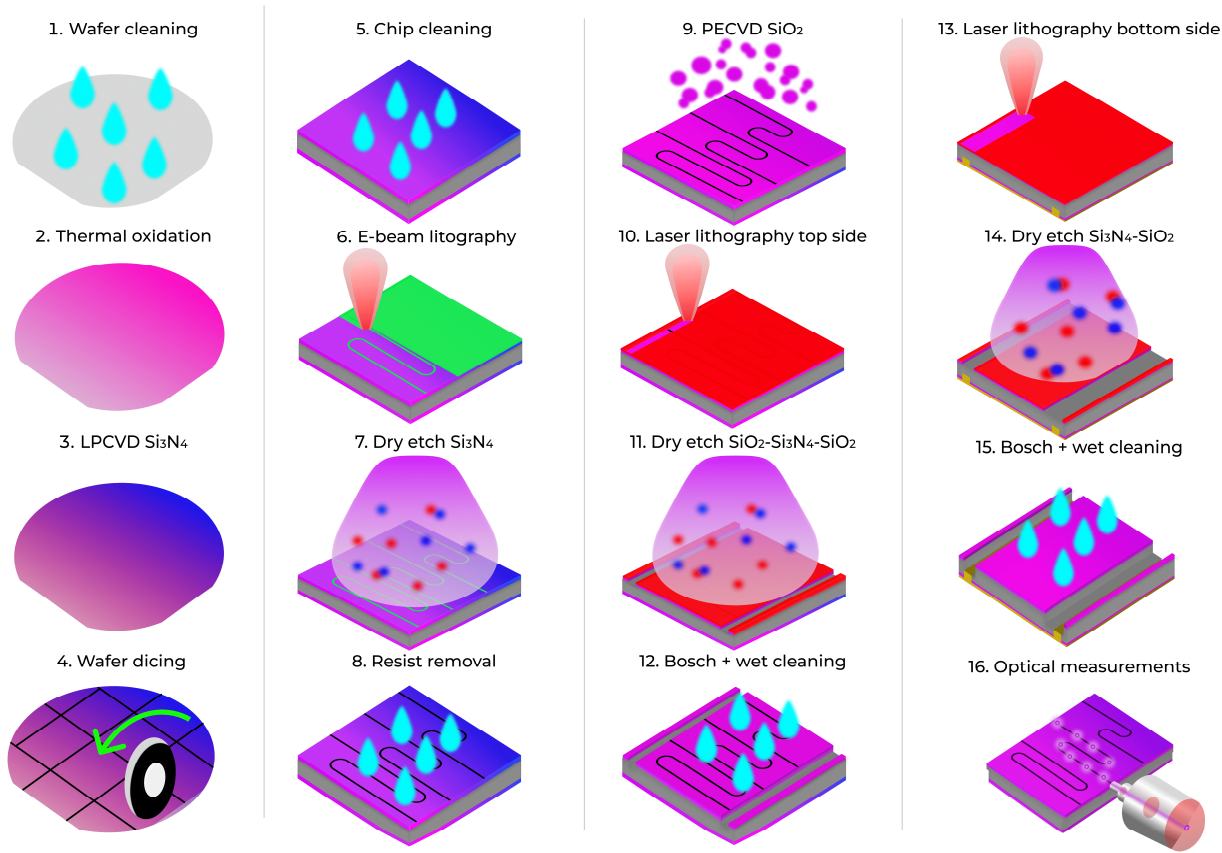

**Figure 3.** Si$_3$N$_4$ PICs with edge coupling fabrication process.

The clear maximum can be seen in Figure 5b. The 88.7° resist sidewall angle corresponds to the following lithography regime: 60 mW laser power, 4 passes, -15 % focus. The found regime was further used in etching technology development.

Etching was performed using the ICP-RIE system. For etching thick layers of silicon dioxide (more than 5 µm) with a vertical profile (90±0.5°) and a smooth edge surface, the CHF$_3$ process gas was selected. During etching process, the CF$_x$ polymer film is deposited on the side walls during the etch process and protects the side surface, simultaneously providing a vertical profile [56,57]. In addition, Ar was added to the plasma to stimulate the etching process.

An increase in the plasma source power in the CHF$_3$/Ar plasma mixture from 1000 W to 2500 W led to an increase in selectivity from 0.9 to 1.3 due to a decrease in the bias voltage. At the same time, the profile angle increased in the positive direction from 88.7° to 92.5° because of the growth of radicals and, as a result, of the process chemical components [58].

An increase in stage power in the range from 50 to 90 W leads to an increase in the bias voltage and an increase in the ionic etching component, which results in a general decrease in selectivity with the local minimum observed at 70 W. With an increase in the process ionic component, the profile angle narrows toward anisotropy (from 88.5° to 86.5°), which, along with the ICP source, is a fairly effective tool for controlling both selectivity and profile angle.

A decrease in the working pressure reduces the total number of collisions within the plasma mixture, thereby increasing the overall direction of the mixture toward the substrate. Through such effects, the profile angle became more negative, and the overall selectivity was reduced because of the presence of more energetic ions in the mixture composition [58].

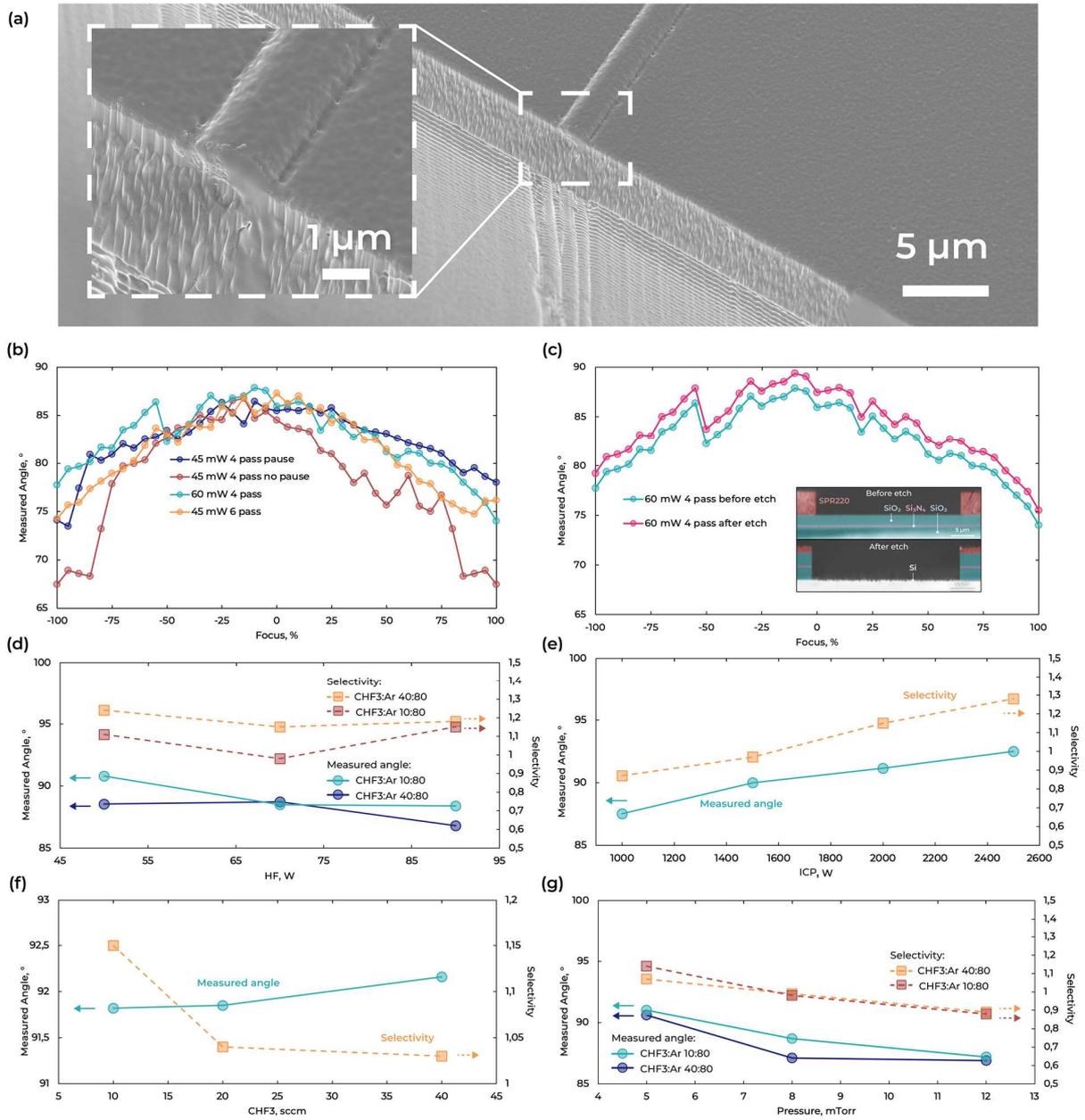

**Figure 4.** SEM image of the fabricated edge coupler (a), dependance of the resist sidewall angle on the laser lithography power and focus value (b), sidewall angle comparison before and after etching (c), $SiO_2$-$Si_3N_4$-$SiO_2$ layers sidewall angle and selectivity dependance on gas mixture (d), $SiO_2$-$Si_3N_4$-$SiO_2$ layers sidewall angle and selectivity dependance on ICP power (e), $SiO_2$-$Si_3N_4$-$SiO_2$ layers sidewall angle and selectivity dependance on $CHF_3$ amount (f) and $SiO_2$-$Si_3N_4$-$SiO_2$ layers sidewall angle and selectivity dependance on chamber pressure (g).

Increasing the amount of the main etching gas $CHF_3$ in the mixture (from 10 sccm to 40 sccm) did not significantly affect the overall process dynamics in terms of the profile angle and remains in the range from 91.85° to 92.15°, which could be considered insignificant. However, a significant increase in the etch rate (from 150 nm/min to 230 nm/min) and a slight decrease in selectivity (from 1.15 to 1.03) were observed. Profile angle preservation with an increase in the etching rate could be associated with preservation of the passivation/etching balance on the channel side walls and simultaneous increase in the ionic and fluorine radicals [59].

Based on the obtained equations, the following etching regime was chosen: stage power 50 W, ICP power 2000 W, pressure 5 mTorr, $CHF_3$ flow 40 sccm, and Ar flow 80 sccm.

**Coupling efficiency improvement.** Thicker bottom and cladding $SiO_2$ layer help to release the vertical mode mismatch and reflection from Si layer. The comparison diagram of value $SiO_2$ thickness and taper type on coupling efficiency is shown on Figure 5.

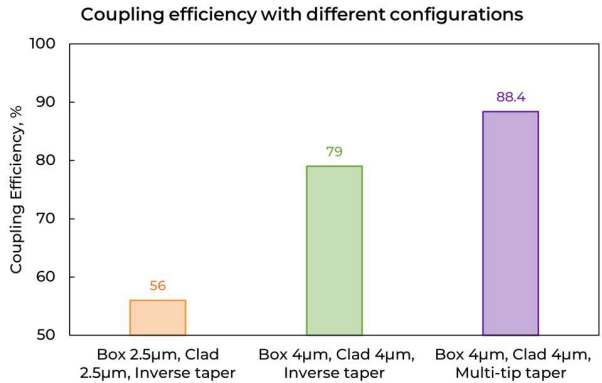

**Figure 5.** Coupling efficiency with different configurations.

To fabricate a photonic integrated circuit with thick oxide layers by deep etching, it is necessary to use a thicker resist layer. The process of forming a resist with a thickness of 9.5 µm and a non-uniformity of 2% was found. For a resist with a thickness of 9.5 µm, a similar work was carried out on selecting the exposure dose and focus as in the previous section. The 89.0° resist sidewall angle corresponds to the following lithography regime: 60 mW laser power, 8 passes, -25 % focus. The figure shows the resist profile angle after the laser lithography step and the profile angle of the optical layers after the etching steps.

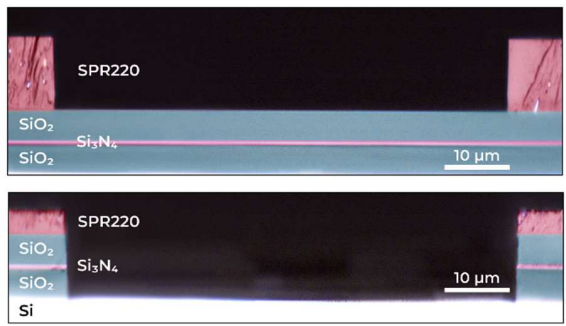

**Figure 6.** Angle profile after laser lithography and after dry etch.

Another approach to reduce coupling losses is to use a multi-tip taper. Improving the geometry of the edge couplers by increasing the tip number often makes improvements to the coupling performance between optical fiber and optical channel on chip [60]. It is known that the mode field diameter of SMF28 fiber is 10 µm, so a wide taper will allow capturing more optical power from the fiber. In addition, the trident-shaped multi-taper will reduce the mismatch of the effective indices of the fiber and the multi-taper.

The proposed multi-taper design is shown in Figure 7. The design consists of three sections (Figure 7): a trident, a transition region, and an adiabatic taper. The tip length, tip start width, connection length and taper length with highest coupling efficiency are calculated by Ansys Lumerical FDTD. The results are shown in Figure 7. The chosen geometry of multi-taper has the following parameters: tip length = 220 um, tip initial width = 100 nm, connection length = 5 um and taper length = 40 um. The calculated and measured coupling losses with such multi-taper were 0.51 dB and 1.5 dB, respectively.

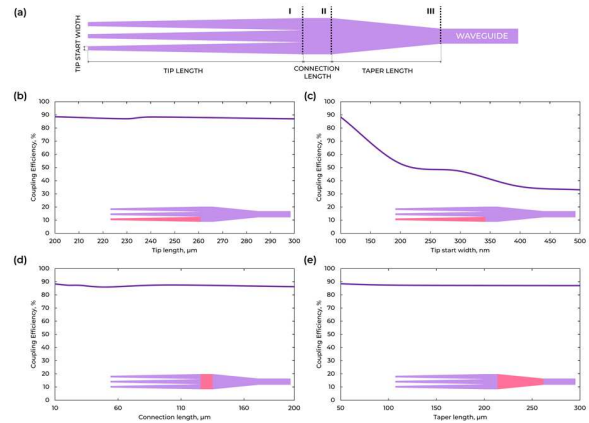

**Figure 7.** Visualization of multi-tip taper with separation on parts (a), Calculation of the tip length for multi-tip taper (b), Calculation of the tip start width for multi-tip taper (c), Calculation of the connection length for multi-tip taper (d), Calculation of the taper length for multi-tip taper (e).

## CHARACTERIZATION

**Cut-back measurement.** To measure the I/O efficiency, an automated assembly system equipped with a 5-nm resolution 12-axis alignment drive, 935 nm and 1550 nm wavelength laser sources, and an optical power meter was used. This system made it possible to measure internal optical losses and coupling efficiency using the output power values of structures of various lengths.

The cut-back propagation loss analysis [61, 62] in the fabricated $Si_3N_4$ waveguides was performed for an integrated circuit. The light was coupled out from the polished FAU through PIC to the power detector. Propagation losses were measured on the test photonic integrated circuits with varying length structures for loss measurement and are shown in Figure 8a.

Figure 8b presents the measured coupling losses for different fiber type and different thickness SiO2 layers. The lowest coupling losses for inverted taper of about -0.81 dB were observed for UHNA-7 fiber. For higher cladding oxide thickness we observed losses minimization due to -0.15 dB for UHNA-7 fiber.

**OFDR measurements.** Reflectometry is widely used in fiber optics to probe the local reflectivity of waveguides and devices with respect to propagation distance [63]. With a point-to-point resolution of about 10 µm and detection sensitivity of −130 dB over 30 m of propagation, coherent optical backscattering reflectometry (OFDR) is a particularly useful technique in characterizing waveguides and devices at the planar scale [64,65]. In the OFDR, a continuous wave laser source is scanned over several terahertz in frequency or, equivalently, several tens of nanometers in wavelength. A larger scan range improves the measurement spatial resolution according to the following expression:

$$D_{min} \cong \frac{c}{2n_g|f_{start} - f_{end}|} = \frac{\lambda_{start}\lambda_{end}}{2n_g|\lambda_{start} - \lambda_{end}|}$$

where $D_{min}$ is the minimum distance between the two data points, c is the speed of light, $n_g$ is the group index, $f_{start}$ and $\lambda_{start}$ are the source frequency and wavelength at the scanning start, respectively, and fend and $\lambda_{end}$ are the source frequency and wavelength at the scanning end [32]. However, all parameters derived from the spatial domain data are then averaged over the measurement scan spectral range. If the parameter spectral dependance is set, the rectangular window function can be applied to data in the spectral domain to narrow the included spectral range. This window could be moved across the measurement's full spectral range, making it possible to extract the parameter at each window position and obtain the parameter spectral dependance from a single OFDR scan. Because narrowing data in the frequency domain decreases the measurement spatial resolution according to Eq. (2), a tradeoff appears between the spectral averaging, which could distort the actual spectral dependance and the measurement accuracy. In this study, a FDTD window width of 10 nm is used to extract the spectrally dependent measurements of the coupling loss propagation loss.

Figure 8d shows the OFDR data from two UHNA-7 fibers coupled to waveguides (10 cm long). The data was not filtered in the spectral domain, and a moving average filter with 100-datapoint or ~1 mm window size was applied in the spatial domain to reduce the backscatter amplitude deviation. Before OFDR scanning, fiber-to-chip coupling was maximized using an OFDR source laser and an optical power meter. The dashed lines in Figure 8d indicate the mean backscatter levels in waveguide. The difference between these two levels of 5.96 dB includes the total return loss between the two fibers, or:

$$RL_{dB}^{total} = 2IL_{dB}^{total} = 2(IL_{dB}^{fiber-to-chip} + IL_{dB}^{propagation} + IL_{dB}^{fiber-to-chip})$$

where $RL_{dB}^{total}$ is the total return loss in dB, total $IL_{dB}$ is the total insertion loss, fiber-to-chip $IL_{dB}$ is the fiber-to-chip insertion loss per facet, and propagation $IL_{dB}$ is the total propagation insertion loss.

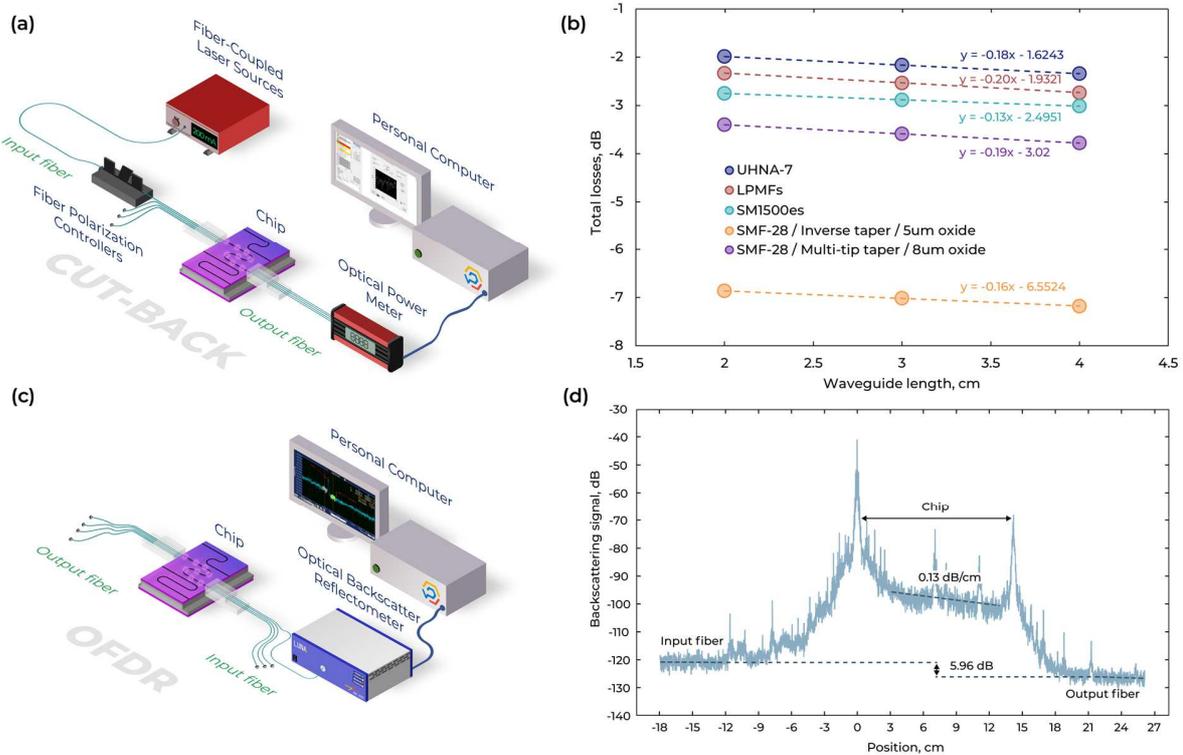

**Figure 8.** Schematic of the measurement setup for cut-back characterization (a); Propagation losses measured by the cut-back calculated: $y = k \times x + b$; $k = propagation\ losses, b = coupling\ losses$ for wavelength 1550 nm (b); Schematic of the measurement setup for OFDR characterization (c) and OFDR characterization along the waveguide. The dashed line shows the linear fit of the waveguide reflections (d).

The measured data is linearly fitted to find propagation losses (Figure 8d). Based on OFDR characterization, the propagation insertion loss is 0.13 dB/cm and fiber-to-chip insertion loss per facet is 0.84 dB.

## CONCLUSION

This study presents a universal strategy for repeatable wafer-scale PIC low-loss edge coupling. For a coupling platform, we present a tutorial on the design, manufacturing, and characterization of lithographically defined optical coupling facets using an ICP dry etching technique. Fabrication process optimization allowed us to obtain the 89 degrees optical grade facet that provides coupling losses lower than 1 dB for a set of optical fibers that agrees well with the simulation results. To reduce propagation insertion losses, fabrication process should include chemical-mechanical polishing and long-term annealing steps. We believe that this research would be useful for a wide PIC community.

AUTHOR INFORMATION

Corresponding Author

**Ilya A. Rodionov** - Shukhov Labs, Quantum Park, Bauman Moscow State Technical University, Moscow 105005, Russia; Dukhov Automatics Research Institute, Moscow 127055, Russia; orcid.org/0000-0002-8931-5142; Email: irodionov@bmstu.ru

Authors

**Sergey A. Avdeev** – Shukhov Labs, Quantum Park, Bauman Moscow State Technical University, Moscow 105005, Russia; Dukhov Automatics Research Institute, Moscow 127055, Russia; orcid.org/0000-0002-7296-367X; Email: avdeevss@bmstu.ru

**Aleksandr S. Baburin** - Shukhov Labs, Quantum Park, Bauman Moscow State Technical University, Moscow 105005, Russia; Dukhov Automatics Research Institute, Moscow 127055, Russia; orcid.org/0000-0003-2806-018X; Email: baburin@bmstu.ru

**Evgeniy V. Sergeev** - Shukhov Labs, Quantum Park, Bauman Moscow State Technical University, Moscow 105005, Russia; orcid.org/0000-0002-2272-2624; Email: sergeev_e@bmstu.ru

**Alexei B. Kramarenko** - Shukhov Labs, Quantum Park, Bauman Moscow State Technical University, Moscow 105005, Russia; orcid.org/0009-0003-2217-8459; Email: kramarenko@bmstu.ru

**Arseniy V. Belyaev** - Shukhov Labs, Quantum Park, Bauman Moscow State Technical University, Moscow 105005, Russia; Email: arsbel99@gmail.com

**Danil V. Kushnev** - Shukhov Labs, Quantum Park, Bauman Moscow State Technical University, Moscow 105005, Russia; Email: dvkushnev@gmail.ru

**Kirill A. Buzaverov** - Shukhov Labs, Quantum Park, Bauman Moscow State Technical University, Moscow 105005, Russia; orcid.org/0000-0002-5550-6880; Email: kirillbuz@bmstu.ru

**Ilya A. Stepanov** - Shukhov Labs, Quantum Park, Bauman Moscow State Technical University, Moscow 105005, Russia; orcid.org/0000-0002-6533-4373; Email: stepanovia@bmstu.ru

**Vladimir V. Echeistov** - Shukhov Labs, Quantum Park, Bauman Moscow State Technical University, Moscow 105005, Russia; orcid.org/0000-0002-2258-9843; Email: Wecheistov@bmstu.ru

**Sergey V. Bukatin** - Shukhov Labs, Quantum Park, Bauman Moscow State Technical University, Moscow 105005, Russia; orcid.org/0009-0006-0557-2659; Email: bukatin@bmstu.ru

**Ali Sh. Amiraslanov** - Shukhov Labs, Quantum Park, Bauman Moscow State Technical University, Moscow 105005, Russia; orcid.org/0009-0003-1744-6420; Email: amiraslanovash@student.bmstu.ru

**Evgeniy S. Lotkov** - Shukhov Labs, Quantum Park, Bauman Moscow State Technical University, Moscow 105005, Russia; orcid.org/0000-0001-5386-7837; Email: lotevg@bmstu.ru

**Dmitry A. Baklykov** - Shukhov Labs, Quantum Park, Bauman Moscow State Technical University, Moscow 105005, Russia; orcid.org/0000-0002-3595-7510; Email: dimabaklykov@bmstu.ru

Author Contributions

Ilya A. Rodionov initiated and directed the work. Sergey S. Avdeev conducted the design, fabrication, measurement and analyzed the experimental data. Aleksandr S. Baburin supervised the project. The manuscript was discussed and corrected by all authors.

Notes

The authors declare no competing financial interest.

## ACKNOWLEDGMENT

Technology was developed and samples were fabricated at Quantum Park (BMSTU Nanofabrication Facility, Shukhov Labs, FMNS REC, ID 74300).